\documentclass[prl,twocolumn]{revtex4}
\usepackage{graphicx}
\usepackage{dcolumn}
\usepackage{bm}

\begin{document}

\title{Symmetry-Based Selective Femtosecond Coherent Control of Transient Two-Photon Absorption}

\author{Andrey Gandman, Leonid Rybak, Michal Bronstein, Naser Shakour,}
\author{Zohar Amitay}%
\email{amitayz@tx.technion.ac.il} %
\affiliation{Schulich Faculty of Chemistry, Technion - Israel Institute of Technology, Haifa 32000, Israel}

\begin{abstract}
We present and implement a new scheme for independent control of
both the final and the transient population utilizing the symmetry
properties of the system.
By proper pulse shaping, utilizing the invariance of the two-photon
absorption to specific phase transformations of the pulse, different
time evolutions of the transient population are photo-induced for a
given (fixed) final state population.
The model system is the Na atom.
The work is conducted in the weak-field regime for which the
transient two-photon excitation is described by second-order
perturbation theory.
One most attractive case is the extended family of third-order chirp
pulses which control the population build-up duration independently
of the final population.
\end{abstract}


\maketitle


The general concept of coherent control is to steer a system towards
a desired outcome by utilizing a manifold of interference pathways
induced by a broad femtosecond pulse
\cite{tannor_kosloff_rice_coh_cont,shapiro_brumer_coh_cont_book,warren_rabitz_coh_cont,
rabitz_vivie_motzkus_kompa_coh_cont}.
Till now, this desired objective was mainly to control the
transition probability to at least one of the given (fixed) final
states.
Such control was experimentally demonstrated in various atomic and
molecular systems \cite{dantus_exp_review1_2,
silberberg_2ph_nonres1_2, dantus_2ph_nonres_molec1_2,
baumert_2ph_nonres, silberberg_2ph_1plus1, girard_2ph_1plus1,
leone_res_nonres_raman_control, leone_cars, silberberg-angular-dist,
silberberg_antiStokes_Raman_spect, wollenhaupt-baumert1_2,
amitay_3ph_2plus1_2,amitay_2ph_inter_field1_2,silberberg-2ph-strong-field,
weinacht-2ph-strong-theo-exp,van-der-Zande-Rb,amitay_selective}.
However, we believe that complete control over the transient as well
as final population is of major importance to any forthcoming
application involving multiphoton processes and fast decoherence or
relaxation rates.
Despite the grate progress in coherent control of state-to-state
transition probability, little work was done in controlling the
transient population build-up.
Most of the experimental advances in this area concentrated on
experimental observation \cite{girard_transient} and control
\cite{girard_transient_fresnel,silberberg_transient} of coherent
transients in a two-level atomic system, using one photon absorbtion
process which does not allow control over the final population of
the system, but only over its transient.
Moreover, several important applications were also demonstrated,
including femtosecond spectral electric field reconstruction
\cite{girard_pulse_reconstruction}, high precision calibration of
pulse-shaping setups \cite{girard_shaper_calibration} and most
importantly, time resolved quantum state holography
\cite{girard_quantum_state_holography}.
In this Letter we present for the first time experimental
observation and full coherent control of transient state population
excited via nonresonant two-photon transition by shaped femtosecond
pulses.
The control scheme extends the frequency-domain picture
\cite{silberberg_2ph_nonres1_2} to describe the transient evolution
of the population build-up.
Then, by a proper pulse shaping, utilizing the invariance of the
two-photon absorption to specific phase transformations of the pulse
\cite{amitay_selective}, different transient evolutions of the
population are photo-induced independently of a desired (fixed)
final state population.
The intuitive theoretical framework is also verified experimentally.
Moreover, extended family of chirp pulses were found to be very
robust in controlling separately the final and transient populations
on a picosecond timescale using combination of frequency and time
domain descriptions.
The work is conducted in the weak-field regime for which the
transient two-photon absorption is described by second-order
perturbation theory, which is the lowest order perturbation for
final state phase control.
The model system is the Na atom.
In general, the essence of the present work is to provide a rational
intuitive way to the full control over the state's population
evolution as well as its final value on both, short and long
timescales.
%


%
We consider an atomic two-photon absorption process from an initial
ground state $\left|g\right>$ to a final excited state
$\left|f\right>$, which is coupled via a manifold of states
$\left|v\right>$ that are far from resonance and have the proper
symmetry.
The light-matter interaction with a shaped temporal electric field
$\varepsilon(t)$ is described by second-order time-dependent
perturbation theory.
\begin{figure} [bp]
\includegraphics[scale=0.25]{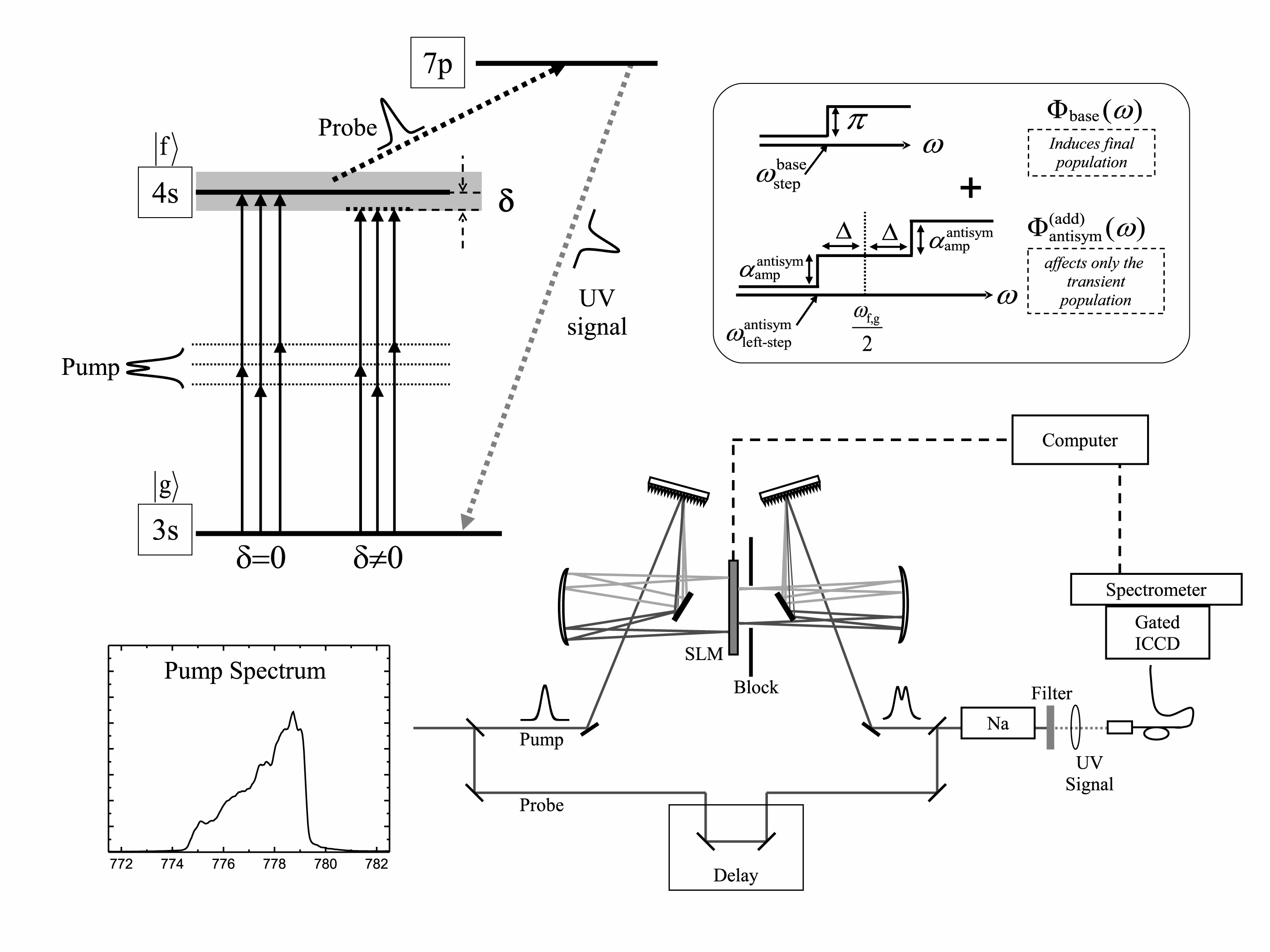}
\caption{\label{fig1} (a) Energy level diagram of the pump-prob
excitation and detection scheme of Na. (b) Representation of the
selective transient coherent control scheme including the base phase
pattern and the anti-symmetric phase additions. (c) Pump pulse
experimental spectra. (d) Outline of the experimental setup. }
\end{figure}
Thus, the time-dependent amplitude $a_{f}(t)$ of state
$\left|f\right\rangle$ at time $t$, is given by
\begin{eqnarray}
\label{eq_af_t}  
 a_{f}(t) =  -\frac{1}{\hbar^{2}}\sum_{v}
   \mu_{fv}\mu_{vg}
   \int_{-\infty}^{t}\int_{-\infty}^{t_1}
    \varepsilon(t_1)\varepsilon(t_2)  \\ \nonumber
    \times \exp[i(\omega_{fv}t_1+\omega_{vg}t_2)]
    dt_1dt_2 ,
\end{eqnarray}
where $\mu_{ij} = \left\langle i \right| \mu \left| j \right\rangle$
is the dipole matrix element between a pair of states and
$\omega_{ij} = (E_{i} - E_{j})/\hbar$ is the corresponding
transition frequency.
Within the frequency-domain framework, the spectral field $E(\omega)
\equiv \left|E(\omega)\right| \exp \left[ i\Phi(\omega) \right]$ is
given as the Fourier transform of $\varepsilon(t)$, with
$\left|E(\omega)\right|$ and $\Phi(\omega)$ being the spectral
amplitude and phase at frequency $\omega$. The transient amplitude
$A_{tr}^{(2)}(t)$ can be written as
\begin{equation}
\label{eq_Atr_wt} %
A_{tr}^{(2)}(t) = \frac{\mu_{fg}^2}{2 \pi i \hbar^2} %
\int_{-\infty}^{\infty}\int_{-\infty}^{\infty}
\frac{E(\omega)E(\omega')}{\omega_{fg}-(\omega+\omega')}
\exp[i(\omega_{fg}-(\omega+\omega')t] ,
\end{equation}
where $\omega_{fg}$ and $\mu_{fg}^{2}$ are, respectively, the
$\left|g\right>$-$\left|f\right>$ transition frequency and effective
non-resonant two-photon coupling.
The contribution to the transient amplitude can be separated into
two channels of excitation:
\begin{equation}
\label{eq_Atr_w} %
A_{tr}^{(2)}(t) = \frac{\mu_{fg}^2}{2 \pi i \hbar^2} %
\left[ A_{on-res}^{(2)} + A_{near-res}^{(2)}(t) \right]  ,
\end{equation}
\begin{equation}
\label{eq_A_res} %
A_{on-res}^{(2)} = i\pi A^{(2)}(\omega_{fg})  ,
\end{equation}
\begin{equation}
\label{eq_A_near_res_t} %
A_{near-res}^{(2)}(t) = -\wp \int_{-\infty}^{\infty}
\frac{1}{\delta} A^{(2)} (\omega_{fg}-\delta) \exp(-i\delta t)
d\delta ,
\end{equation}
with
\begin{equation}
\label{eq_A2_omega} %
A^{(2)}(\Omega)  =  \int_{-\infty}^{\infty} E(\omega)
E(\Omega-\omega)d\omega ,
\end{equation}

The above frequency domain description allows intuitively to
identify the interfering pathways that coherently contribute to the
transient two-photon transition amplitude (see Fig.~\ref{fig1}).
The time independent amplitude $A_{on-res}^{(2)}$ coherently
interferes all the possible two-photon pathways from
$\left|g\right\rangle$ to $\left|f\right\rangle$ $(\delta=0)$, i.e.,
coherently integrates over all their corresponding amplitudes. Each
such pathway is composed of two absorbed photons of frequencies
$\omega$ and $\omega_{fg}-\omega$.
As opposed to the time dependent amplitude $A_{near-res}^{(2)}(t)$
which coherently interferes all the possible near-resonance
two-photon pathways having a corresponding detuning $(\delta\ne0)$.
It involves a non-resonant absorption of two photons with a
two-photon transition frequency $\omega_{fg}-\delta$ with a
$1/\delta$ amplitude weighting and a time dependent phase term
$\exp(-i\delta t)$.
The on-resonant pathways are excluded from $A_{near-res}^{(2)}(t)$
by the Cauchy's principal value operator $\wp$.
The different amplitudes are expressed using the parameterized
amplitude $A^{(2)}(\Omega)$ interfering all the possible two-photon
pathways with transition frequency $\Omega$: $A_{on-res}^{(2)}$ is
proportional to $A^{(2)}(\omega_{fg})$, while
$A_{near-res}^{(2)}(t)$ integrates all $A^{(2)}(\omega_{fg}-\delta)$
with $\delta \ne 0$.
The transient two-photon transition probability is given by
$P_{tr}^{(2)}(t) = \left|A_{on-res}^{(2)}+
A_{near-res}^{(2)}(t)\right|^2$.
In agrement with previous studies, by evaluating
Eq.~(\ref{eq_Atr_w}) for time $t\rightarrow\infty$ (i.e., after the
pulse is over), one obtains
$A_{near-res}^{(2)}(t\rightarrow\infty)=A_{on-res}^{(2)}$ for finite
amplitude of state $|f\rangle$, and the transition probability
becomes $P_{tr}^{(2)}(t\rightarrow\infty) =
\left|2A_{on-res}^{(2)}\right|^2$.
%


%
Utilizing the symmetry property of $A^{(2)}(\omega_{fg})$ that is
invariant to any antisymmetric phase transformation, the two
channels of excitation (i.e., the on-resonance time independent
channel $A_{on-res}^{(2)}$ and the near-resonance transient channel
$A_{near-res}^{(2)}(t)$) can be selectively controlled.
Any phase $\Phi_{\textrm{antisym}}^{(\textrm{add})}(\omega)$ that is
antisymmetric around $\omega_{fg}/2$ keeps the value of
$A^{(2)}(\omega_{fg})$ unchanged, while it generally changes the
value of $A^{(2)}(\omega_{fg}-\delta)$ for $\delta\ne0$. Hence, it
alters $A_{near-res}^{(2)}(t)$ while keeping $A_{on-res}^{(2)}$
invariant.
The effect of antisymmetric phase addition is explained in details
in \cite{amitay_selective}.
$A_{on-res}^{(2)}$, on the other hand, can be controlled from a zero
till to its maximal value (TL-level) using a simple $\pi$-step
spectral base phase $\Phi_{\textrm{base}}(\omega)$.
The total phase pattern
$\Phi(\omega)=\Phi_{\textrm{base}}(\omega)+\Phi_{\textrm{antisym}}^{(\textrm{add})}(\omega)$
is presented schematically in Fig~\ref{fig1}.
%


%
The physical model system of the study is the sodium (Na) atom, with
the $3s$ ground state as $\left|g\right\rangle$ and the $4s$ final
state state as $\left|f\right\rangle$, (see Fig.~\ref{fig1}).
The transition frequency $\omega_{fg} \equiv \omega_{4s,3s} =
25740$~cm$^{-1}$ corresponds to two 777-nm photons.
The $3s$-$4s$ two-photon coupling is nonresonant via the $p$ states.
The Na atom interacts with phase-shaped linearly-polarized
femtosecond pulses having a one-side blocked at 779~nm
(12837~cm$^{-1}$) Gaussian intensity spectrum centered around 790~nm
(12658~cm$^{-1}$) with 4~nm (66-cm$^{-1}$) effective bandwidth
($\sim$200-fs TL duration), (see Fig.~\ref{fig1}).
%


%
The sodium vapor in a heated cell is irradiated with such laser
pulses, after they undergo shaping in an optical setup incorporating
a pixelated liquid-crystal spatial light phase modulator
\cite{pulse_shaping}. The effective spectral shaping resolution is
$\delta\omega_{shaping}$=2.05~cm$^{-1}$ (0.125~nm) per pixel.
The peak intensity of the TL pulse is below 10$^{9}$~W/cm$^{2}$,
corresponding to the weak-field regime.
To demonstrate the $\left|f\right\rangle$ state transient population
$P^{(2)}_{tr}(t)$, the $4s$ state is probed to the $7p$ state by
weak unshaped linearly-polarized femtosecond pulses having a
Gaussian intensity spectrum centered around 790~nm (12658~cm$^{-1}$)
with 12-nm (300-cm$^{-1}$) bandwidth ($\sim$80-fs TL duration), at
time delay $\tau_p$. The experimental setup is presented
schematically in Fig.~\ref{fig1}.
The $4s$-$7p$ transition frequency $\omega_{7p,4s} =
12801$~cm$^{-1}$ corresponding to a one 781.2~nm photon is included
in the probe pulse however, it is excluded in the pump pulse by
blocking this part of the spectrum at the Fourier plane of the pulse
shaper.
The probe beam, which remains a TL pulse, is spatially overlapped
with the pump beam in the heated sodium cell.
We evaluate the transient population $P^{(2)}_{tr}(t)$ of the $4s$
state by probing the third harmonic generation signal at the
$7p$-$3s$ transition frequency $\omega_{7p,3s} = 28541$~cm$^{-1}$
corresponding to a one 259.5~nm UV-photon \cite{amitay_UV}.
We assume that the probe pulse is sufficiently weak and short
compared to the shaped pump pulse.
The UV signal is measured within the beam propagation direction
using a spectrometer coupled to a time-gated camera system.
%

\begin{figure} [t]
\includegraphics[scale=0.32]{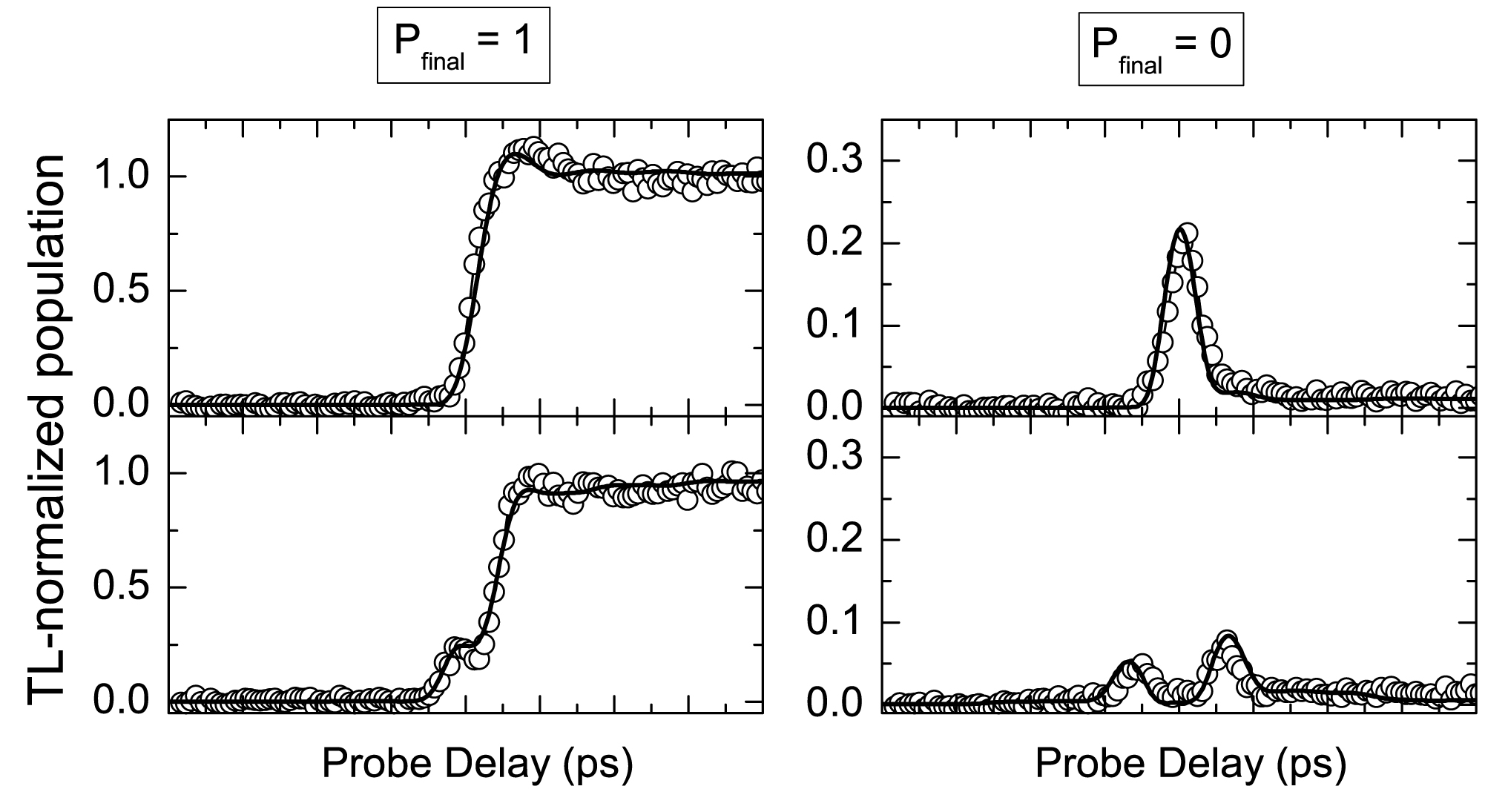}
\caption{\label{fig2} (a)-(d) Experimental (circles) and
numerical-theoretical (lines) results for coherent control of
transient two-photon absorption. Panels (a) and (b) show different
transient population build-ups resulting in the same, TL-level final
population. Panels (c) and (d) show different transient population
resulting in zero final population, (see text). All the data is
normalized by $P^{(2)}_{TL}(t\rightarrow\infty)$ excited by the
transform-limited pulse. The data is presented as a function of the
probe pulse delay.}
\end{figure}

%
Figure \ref{fig2} demonstrates experimentally the transient
selective coherent control strategy.
It presents experimental (circles) and numerical-theoretical (lines)
results (Eqs.~\ref{eq_Atr_w}-\ref{eq_A2_omega}) for coherent control
of transient two-photon absorption.
All the data is normalized by the final population
$P^{(2)}_{TL}(t\rightarrow\infty)$ excited by the transform-limited
pulse.
The data is presented as a function of the probe pulse delay.
Panels (a) and (b) show different transient population build-ups
resulting in the same TL-level final population.
While panel (a) shows the transient population for a TL pulse, i.e.,
both $\Phi_{\textrm{base}}(\omega)=0$ and
$\Phi_{\textrm{antisym}}^{(\textrm{add})}(\omega)=0$ for any
$\omega$, panel (b) presents the transient population of a modified
pulse based on the symmetry properties of the system for which
$\Phi_{\textrm{base}}(\omega)=0$ remains zero for any $\omega$,
however $\Phi_{\textrm{antisym}}^{(\textrm{add})}(\omega)$ is not.
It has a shape of an antisymmetric step (see Fig.~\ref{fig1}) with
$\Delta=0.55~nm$ and $\alpha^{antisym}_{amp}=\pi$.
Although the final population is not affected by this phase addition
the transient population build-up is very different as compared to
the TL pulse.
As long as for the former (a), the population transfer is a smooth
line from zero population before the pulse to TL-level at the end of
the pulse, for the later (b), it is much longer in time and has a
kink in the vicinity of zero probe delay.
%

As another example Fig.~\ref{fig2} (c) and (d) present the same
scenario of selective transient coherent control for zero final
population.
Panel (c) shows the transient population evolution which eventually
results in zero population transfer, i.e., so called a
\emph{dark}-pulse \cite{silberberg_2ph_nonres1_2}.
It utilizes non-zero base phase with a $\pi$-step at
$\omega^{\pi,dark}_{base}=778.2~nm$ which assures zero final
population. The antisymmetric addition
$\Phi_{\textrm{antisym}}^{(\textrm{add})}(\omega)=0$ remains zero
for any $\omega$.
First, the transient population is gradually accumulated in the
final state until it reaches a value of $~20\%$ of the TL-level when
the probe delay is zero.
Then, all the accumulated population is transferred back to the
ground state.
In panel (c), while, the base phase with a $\pi$-step at
$\omega^{\pi,dark}_{base}=772.2~nm$ assures a zero final population,
the antisymmetric addition
$\Phi_{\textrm{antisym}}^{(\textrm{add})}(\omega)$  with
$\Delta=0.55~nm$ and $\alpha^{antisym}_{amp}=\pi$ alters completely
the transient population evolution without changing the zero final
population.
It has two peaks across the evolution with a deep around zero probe
delay, and as in the case of TL-level final population, in panels
(a) and (b), the modified pulse induces much longer evolution of the
transient population.
It worth mentioning that all the experimental results are in
excellent agrement with the theoretical calculations.
%


\begin{figure} [t]
\includegraphics[scale=0.2]{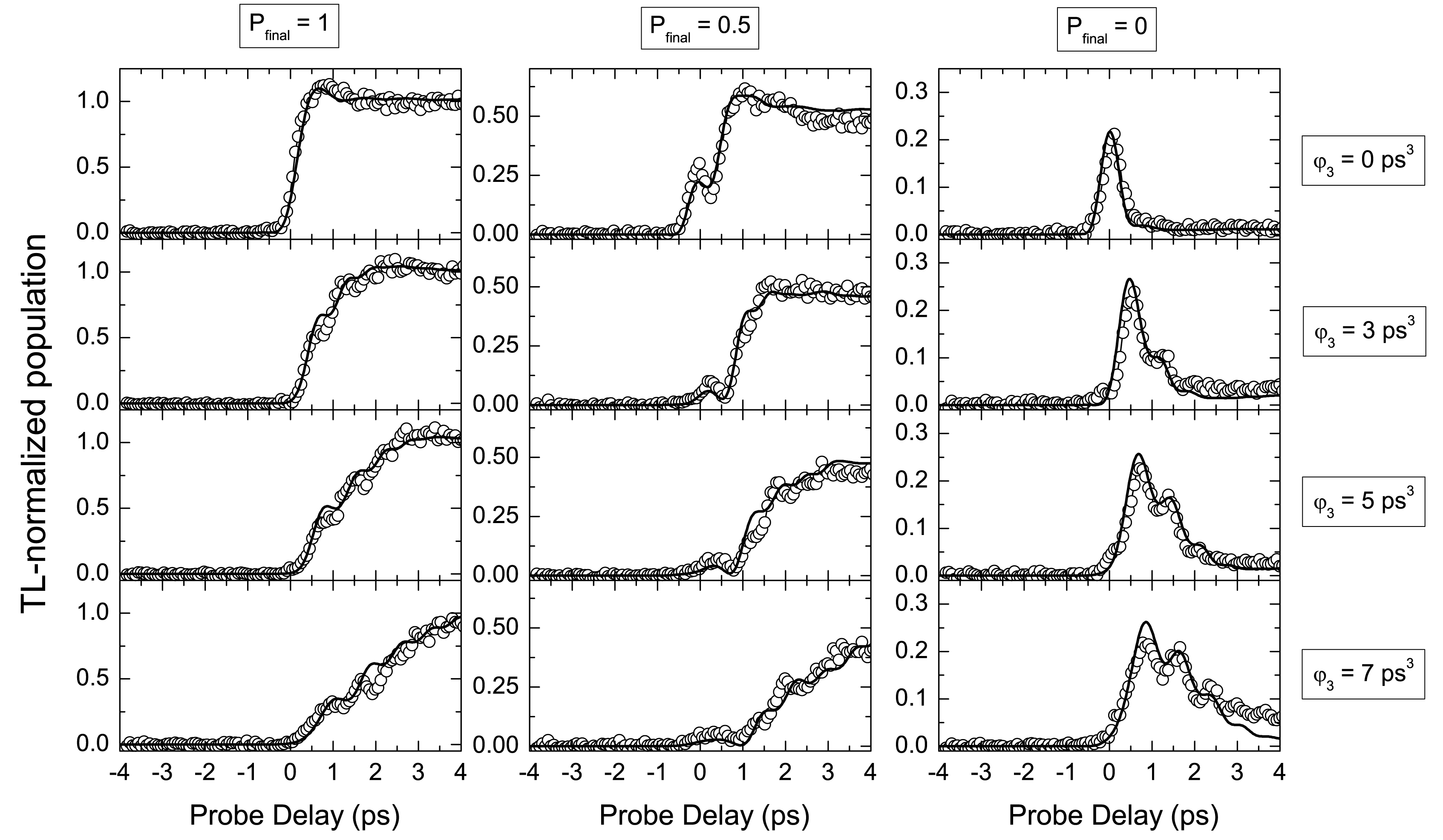}
\caption{\label{fig3} Experimental (circles) and
numerical-theoretical (lines) results for symmetry based independent
control of the duration of the transient population evolution
regardless of the desired final population. The left column presents
the results for TL-level final population with increasing cubic
spectral phase from $\varphi_3=0$ for the first row,
$\varphi_3=3$~ps$^3$ for the second row, $\varphi_3=5$~ps$^3$ for
the third, and $\varphi_3=7$~ps$^3$ for the forth. The same cubic
phase additions were applied in the middle, and right rows
respectively for half of the TL-level final population and zero
final population. The data is presented as a function of the probe
pulse delay. }
\end{figure}

The presented symmetry based control scheme can be applied to
independent control of the duration of the transient population
evolution regardless of the desired final population.
The experimental (circles) together with the theoretical (line)
results are presented in Fig.~\ref{fig3}.
In this results the final population is controlled using
$\Phi_{\textrm{base}}(\omega)$ in a form of a simple $\pi$-step by
setting the step at different position $\omega^{base}_{step}$.
The duration of the transient population evolution is controlled
using quadratic chirp spectral phase addition
$\Phi_{\textrm{antisym}}^{(\textrm{add})}(\omega)=\varphi_3\left(\omega-\omega_{fg}/2\right)^3$,
where $\varphi_3$ is a third-order coefficient.
This phase addition is antisymmetric around the half of the
two-photon transition frequency $\omega_{fg}/2$. Being so, it will
not influence the final population set by the base phase.
The left column in Fig.~\ref{fig3} presents the results for TL-level
final population with increasing cubic spectral phase from
$\varphi_3=0$ for the first row, $\varphi_3=3$~ps$^3$ for the second
row, $\varphi_3=5$~ps$^3$ for the third, and $\varphi_3=7$~ps$^3$
for the forth.
The same cubic phase additions were applied in the middle, and right
rows respectively for half of the TL-level final population and zero
final population.
As Fig.~\ref{fig3} clearly indicates, the duration of the transient
population evolution is highly correlated to the magnitude of the
quadratic chirp addition, regardless of the final population.
As we increase the magnitude of the quadratic chirp, the duration of
transients also increase.
This transient elongation is very intuitive as it is well known that
quadratic chirp produce beats in the intensity vs. time, causing
oscillations after the main pulse.
Those oscillations become more and more pronounce as we incense the
coefficient $\varphi_3$, effectively elongating the pulse.
It worth mentioning that linear chirp addition is not applicable for
the current scenario as it is symmetric around $\omega_{fg}/2$ so it
will alter the final population together with the transient
duration.
Although, in general all the experimental results are in excellent
agrement with the theoretical calculations, some discrepancies that
do appear, can be accounted for finite pulse shaper resolution
\cite{pulse_shaping} and the high sensitivity of the signal
\cite{girard_shaper_calibration}. The effect is clearly more
pronounced for high chirp values as we get close to the sampling
limitation of the shaper.
%


\begin{figure} [t]
\includegraphics[scale=0.25]{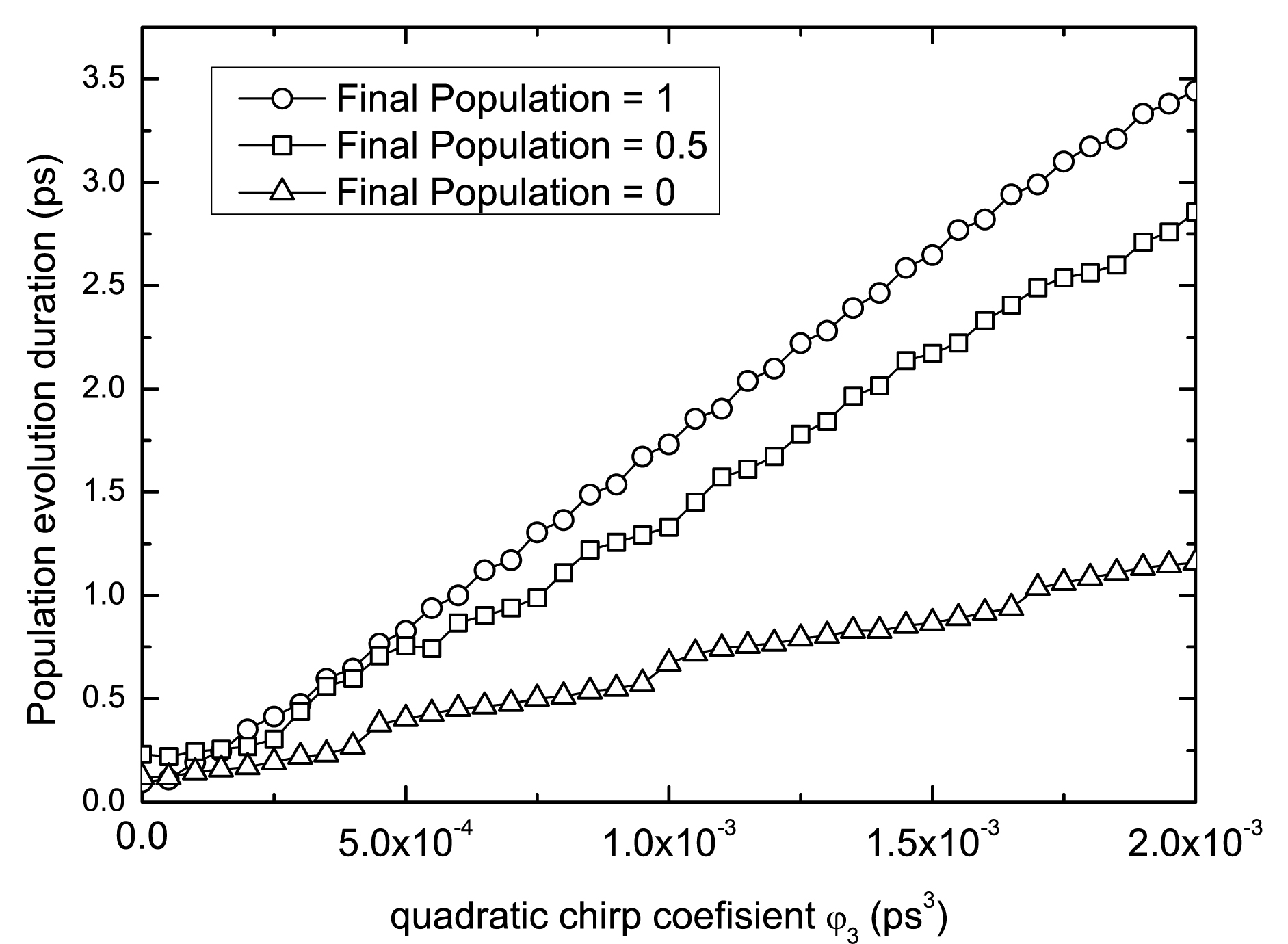}
\caption{\label{fig4} Duration population evolution control.
Calculated results show the duration of transient duration as a
function of quadratic chirp coefficient $\varphi_3$ for different
final populations.}
\end{figure}

To emphasize the robustness of the selective transient coherent
control which allows independent control of transient population
evolution for a given final population, duration of such evolution
is presented as a function of the quadratic chirp coefficient
$\varphi_3$ for different final populations.
The calculated data for a Gaussian pulse corresponding to 100~fs
unshaped duration is presented in Fig.~\ref{fig4}.
The transient duration is calculated by looking at $5\%$ (relative
to TL-level) deviation of the population from the initial and final
levels.
The results clearly indicate the possibility of transient duration
control independent of the final population.
The discrepancies between the curves for deferent final levels are
accounted for the duration calculation procedure which does not take
into account the relativity of the final population.
The common feature for all the curves in Fig.~\ref{fig4} appear to
have little correlation for small values of the chirp and a very
pronounced linear dependence for high chirp values.
As for the first region, both the $\Phi_{\textrm{base}}(\omega)$ in
a form of a simple $\pi$-step and the
$\Phi_{\textrm{antisym}}^{(\textrm{add})}(\omega)$ have similar
relative effect on the pulse duration, thus influencing the
transient duration in the same manner.
As for the linear region, the $\pi$-step influence on the pulse
duration is negligible relative to the applied chirp, thus resulting
in little impact on the duration of the transient, which is
controlled solely by chirp magnitude.
Despite its reduced correlation to the duration of the transient
evolution, the basic part $\Phi_{\textrm{base}}(\omega)$ of the
applied phase is vital for control of the final population by
setting the proper interference pattern.
Such combination of frequency domain picture given in
Eqs.~\ref{eq_Atr_w}-\ref{eq_A2_omega} with the time domain
perspective gives a simple intuitive way to draw the control
scenario for independent selective control of both the transient and
final populations.
While the frequency domain formulation allows us to identify the
interfering pathways selectively contributing to the final and
transient populations, the time domain picture sets the general
limits on the duration of the transient evolution.
%


In conclusion, full selective coherent control of transient
population is presented for the first time in which both transient
and final populations are controlled independently.
The control scheme is very robust, intuitive and general, and can be
applied to any system, once its symmetry properties are identified.
The duration of the transient build-up can be controlled from 100~fs
to several picoseconds for any desired final population.
Moreover, the presented control scheme can be extended to basically
any structure of transient evolution by employing a semirational
however automatic phase optimization based on closed-loop learning
algorithms
\cite{rabitz_feedback_learning_idea,gerber_feedback_control_review,levis_feedback_cont,motzkus_bio1_feedback_cont}.



\newpage

\end{document}